\documentclass[lettersize,journal]{IEEEtran}
\usepackage{amsmath,amsfonts}
\usepackage{algorithmic}
\usepackage{algorithm}
\usepackage{array}
\usepackage[caption=false,font=normalsize,labelfont=sf,textfont=sf]{subfig}
\usepackage{textcomp}
\usepackage{stfloats}
\usepackage{url}
\usepackage{verbatim}
\usepackage{graphicx}
\usepackage{cite}
\usepackage{booktabs}
\usepackage{multirow}
\usepackage{color}
\usepackage{xcolor}
\definecolor{lightred}{rgb}{1.0, 0.5, 0.5} 

\begin{document}

% \title{Reliability Matters: Fine-grained Multiple Speech Attribute Control for Speaker-Independent Speech Emotion Recognition}
\title{MSAC-SERNet: A Reliable Unified Framework for Speaker-Independent Speech Emotion Recognition}

\author{Yu Pan, Yuguang Yang, Yuheng Huang, Jixun Yao, \\ Jingjing Yin, Yanni Hu,  Heng Lu, Lei Ma, Jianjun Zhao
\thanks{\textbullet{} Yu Pan and Jiajun Zhao are with Kyushu University}% <-this % stops a space
\thanks{\textbullet{} Yuguang Yang, Jixun Yao, Jingjing Yin, Yanni Hu, Heng Lu are with Shanghai Ximalaya Technology Co Ltd}
\thanks{\textbullet{} Yuheng Huang and Lei Ma are with University of Alberta}
}
% \author{Yu Pan, Yuguang Yang, Yuheng Huang, Jixun Yao, Jingjing Yin, Yanni Hu, ~\IEEEmembership{Member,~IEEE,} \\ Heng Lu, Lei Ma, Jianjun Zhao, ~\IEEEmembership{Senior Member,~IEEE}
        % <-this % stops a space

%\thanks{Manuscript received April 19, 2021; revised August 16, 2021.}}

\markboth{Journal of \LaTeX\ Class Files,~Vol.~14, No.~8, August~2021}%
{Shell \MakeLowercase{\textit{et al.}}: A Sample Article Using IEEEtran.cls for IEEE Journals}

\maketitle

\begin{abstract}
Despite notable progress, speech emotion recognition (SER) remains challenging due to the intricate and ambiguous nature of speech emotion, particularly in wild world. While current studies primarily focus on recognition and generalization abilities, our research pioneers an investigation into the reliability of SER methods in the presence of semantic data shifts and explores how to exert fine-grained control over various attributes inherent in speech signals to enhance speech emotion modeling. In this paper, we first introduce MSAC-SERNet, a novel unified SER framework capable of simultaneously handling both single-corpus and cross-corpus SER. Specifically, concentrating exclusively on the speech emotion attribute, a novel CNN-based SER model is presented to extract discriminative emotional representations, guided by additive margin softmax loss. Considering information overlap between various speech attributes, we propose a novel learning paradigm based on correlations of different speech attributes, termed Multiple Speech Attribute Control (MSAC), which empowers the proposed SER model to simultaneously capture fine-grained emotion-related features while mitigating the negative impact of emotion-agnostic representations. Furthermore, we make a first attempt to examine the reliability of the MSAC-SERNet framework using out-of-distribution detection methods. Experiments on both single-corpus and cross-corpus SER scenarios indicate that MSAC-SERNet not only consistently outperforms the baseline in all aspects, but achieves superior performance compared to state-of-the-art SER approaches. 
\end{abstract}


\section{Introduction}
\IEEEPARstart{H}{uman} emotion, intricately woven from the threads of personal connections, environmental stimuli, and cognitive processes, stands as a captivating and complex facet of the human experience \cite{mishra2023chirplet}. As a powerful catalyst, emotions not only steer human behavior but also weave their essence into the fabric of daily interactions and professional endeavors \cite{zao2014time}. It is within this nuanced context that the fusion of emotion and technology takes center stage, giving rise to the significant field of speech emotion recognition (SER), of which primary goal is to automatically discern emotional states within spoken communication.
Given the potential to profoundly impact both human well-being and technological advancements, SER plays a crucial role in diverse practical applications, such as human machine interaction \cite{zhang2017speech}, healthcare \cite{latif2020speech}, education \cite{yadegaridehkordi2019affective}, business \cite{8960433}, and so forth.

In the realm of SER, diverse speech signals from speakers of different regions  and recording environments usually exhibit distinct characteristics, yet there exists implicit similar emotional information within the same emotion category, as pointed out in \cite{wen2022ctl}. 
Consequently, the fundamental challenge in SER lies in acknowledging the inherent variability within signals while identifying the intrinsic emotional representations in these information sources, aiming to construct robust models for prediction.

Under such scenarios, how to train an effective and robust classifier for SER based on these signals has been a subject of research for over two decades. Early studies usually focus on single-corpus tasks, where models were developed and evaluated within the confines of a singular emotion corpus.
These approaches emphasize improving the discrimination of hand-crafted features \cite{schmitt2016border,eyben2015geneva,akccay2020speech}, which are then input into conventional machine learning algorithms like support vector machine (SVM) \cite{mower2010framework,schuller2005speaker}, decision trees \cite{schuller2005speaker,lee2011emotion}, or Bayesian classifiers \cite{Kwon2003EmotionRB}.
In recent years, with the remarkable development of machine learning and deep learning technology, deep neural network (DNN) based methods, such as Long Short-term Memory (LSTM) \cite{zhang2019attention,zhao2019speech,liu2023dual}, Recurrent Neural Network (RNN) \cite{chen20183,mirsamadi2017automatic}, Convolution Neural Network (CNN) \cite{9746679,ye2023temporal,latif2022self}, and others \cite{huang2019feature,guizzo2023learning,pan2023gemo,chatziagapi2019data,yi2023exploring}, have become the mainstream approach in SER. Benefiting from the robust feature learning capability of DNNs, the performance of these methods has undergone a significant enhancement compared to traditional approaches.
Afterwards, researchers progressively shifted their attention towards cross-corpus SER tasks, which involve training SER models using one or more public datasets and then evaluating their recognition capabilities and generalization performance with unseen emotional corpora.
These methods often prioritize paradigmatic design, typically integrating the comprehensive utilization of diverse speech attributes within the signal.
For instance, Xiao \emph{et al} \cite{xiao2020learning} presented a generalized domain adversarial neural network (GDANN) and a class-aligned GDANN to learn domain-invariant representations for robust emotion recognition. 
Latif \emph{et al} \cite{latif2022self} introduced an adversarial dual discriminator (ADDi) network and a self-supervised learning based ADDi network to address the performance degradation of the cross-corpus and cross-language SER. 
Yu \emph{et al} \cite{pan2023gemo} advocated a gender-attribute-augmented contrastive language-audio pretraining SER framework with the guidance of natural language supervisions to identify emotions.

Nevertheless, notwithstanding impressive advancements, these approaches still have several limitations, and the SER task remains open and challenging.
\begin{itemize}
\item \emph{First}, there is limited research attention directed towards simultaneously addressing single-corpus and cross-corpus SER tasks \cite{wen2022ctlmtnet}, whereas the demand for such capabilities is imperative in complex real-world applications.
\item \emph{Second}, despite numerous recent studies \cite{hh2018domain,zhou2019transferable,xiao2020learning,ahn2021cross,latif2022self,li2019improved,parry22_interspeech,pan2023gemo} aiming to enhance the SER performance through the incorporation of various additional speech attributes (gender, speaker, language, etc) via adaptive or adversarial domain adaptation methods, correlations among different speech attributes are often overlooked.
For instance, the commonly used voiceprint/speaker attribute based gradient reversal control method, while partially mitigating the speaker bias issue in SER \cite{schuller2005speaker,gat2022speaker,lu2022domain}, also impairs gender attribute information in the speech signal. 
However, according to many literature \cite{li2019improved,parry22_interspeech,pan2023gemo}, gender attributes undoubtedly provide useful information for speech emotion classification. 
\item \emph{Third}, in comparison to other domains like computer vision (CV) \cite{pan2021lightweight,zong2023detrs,tian2020fcos}, natural language processing (NLP) \cite{wang2022deepstruct,le2022perturbations,tao2022compression}, or even tasks within speech processing such as automatic speech recognition \cite{yang2022lmec,gulati2020conformer,yang2023hybridformer}, it is noteworthy that there is a scarcity of publicly available benchmarks for SER, with the majority being of a relatively modest scale.
In this context, the trained SER methods will inevitably encounter unseen or unfamiliar data when deployed in the open world, thus placing high demands on the reliability of the SER methods. However, current SER field largely focuses on the approaches' recognition and generalization performance, while ignoring the reliability performance which is of great importance in real-world applications, such as medical diagnosis, legal assistance, security monitoring, and so on.
\end{itemize}

\begin{figure*}
\centering
	\includegraphics[height=13.5cm,width=!]{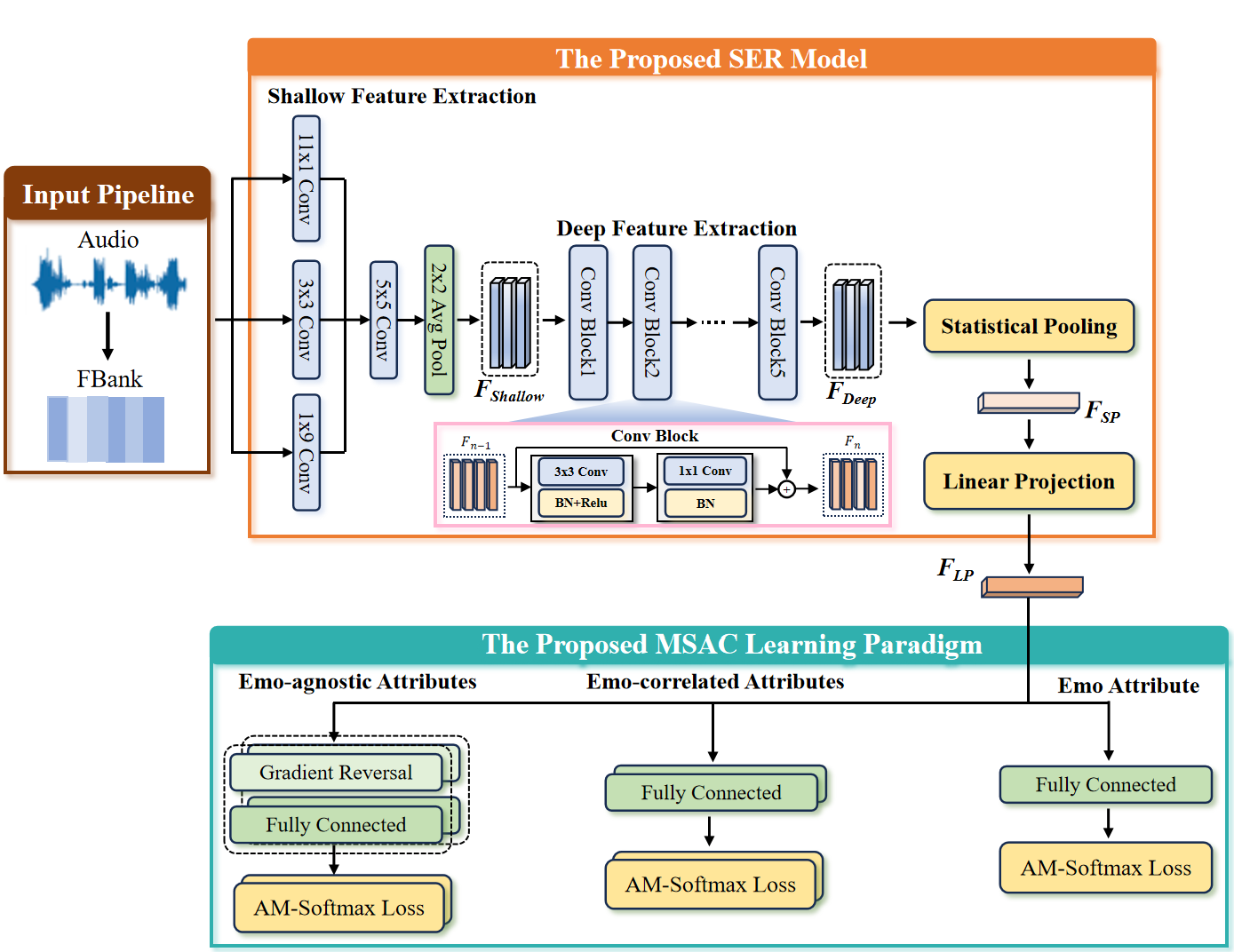}
	\caption{\label{fig:fig1} Overall architecture of the proposed unified SER workflow MSAC-SERNet, including three parts: (1) Input pipeline. (2) The proposed SER mdoel. (3) The proposed MSAC learning paradigm.}
\end{figure*}

Hence in this study, we pioneer an investigation into the reliability of SER methods in the presence of semantic data shifts, and introduce a novel unified SER workflow termed MSAC-SERNet that is able to simultaneously address both single-corpus and cross-corpus SER tasks. Our primary focus is on intricately modeling speech emotion by exercising precise control over diverse speech attributes.
In summary, the main contributions of this paper are four-fold.
\begin{itemize}
\item \emph{First}, from the perspective of merely modeling speech emotion, a novel convolution neural network (CNN) based SER model is constructed, whose architecture is shown in Fig. \ref{fig:fig1}. 
% \yuheng{Perhaps designing modules for feature extraction can be seen as a different contribution to loss design?}
% \panyu{May not understand. Could you explain it more detailed?}
To ensure the discriminability of the acquired features, we design several modules such as shallow feature extraction with different receptive fields, deep feature extraction, and aggregation pooling. Besides, we adopt additive margin softmax (AM-Softmax) \cite{wang2018additive} loss to train the proposed SER model end-to-end, so as to alleviate issues of inter-class similarity and intra-class difference of emotion categories.
\item \emph{Second}, considering the redundancy of various attributes in speech signals, we propose MSAC (multiple speech attribute control), a novel learning paradigm that can explicitly control different speech attributes, enabling the proposed SER model to be less affected by emotion-agnostic speech attributes and better capture nuanced emotion-associated representations. 
In this way, our proposed unified SER workflow MSAC-SERNet could enhance its overall performance.
\item \emph{Third}, four state-of-the-art (SOTA) out-of-distribution (OOD) detection method, i.e., ODIN \cite{ODIN}, Ma\_Distance \cite{MADIS}, MaxLogit \cite{hendrycks22a}, and ReACT \cite{sun2021react} are employed to examine the reliability of our proposed SER models under both single-corpus and cross-corpus SER settings. 
Besides, we also propose a simple OOD detection method rODIN (reverse ODIN), which achieves the best results under both single and cross corpus SER settings. 
To the best of our knowledge, this is the first work to evaluate and analyze the reliability performance of SER methods in the presence of semantic data shifts.
\item \emph{Fourth}, we conduct comprehensive experiments on both single-corpus and cross-corpus SER tasks, employing six public speech emotion corpora. The results illustrate that our proposed MSAC-SERNet attains superior performance in comparison to SOTA SER approaches across both single-corpus and cross-corpus SER scenarios. Notably, it consistently outperforms the baseline model in all facets within the speaker-independent mode.
\end{itemize}

The rest of this work is arranged as the following. In Section \ref{sec:Related Work}, we provide the overview of related work. Section \ref{sec:METHODOLOGY} describes the details of the proposed MSAC-SERNet which mainly consists of three parts, as depicted in Fig \ref{fig:fig1}. In Section \ref{sec:EXPERIMENTS}, we introduce the corresponding experimental settings, results, and analysis of the single-corpus and cross-corpus SER tasks. Section \ref{sec:CONCLUSIONS} draws the conclusions.

\section{Related Work}
\label{sec:Related Work}

In this section, we briefly present the related work for the speaker-dependent and speaker-independent SER, adversarial domain adaption, and out-of-distribution (OOD) detection for better legibility.

\subsection{Speaker-dependent and Speaker-independent SER}

Current SER field can be categorized into speaker-dependent SER \cite{dahake2016speaker,zhao2019speech,deng2017semisupervised} and speaker-independent SER \cite{chen20183,gat2022speaker,wagner2023dawn,pan2023gemo} on the basis of whether the same speaker is utilized during both the training and testing phases.

In general, the former relies on training and testing data from the same or partially overlapping set of speakers, emphasizing the recognition of individualized emotional nuances. 
In \cite{dahake2016speaker}, features extracted using a hybrid of pitch, formants, MFCC, and their statistical parameters were input into an SVM classifier.
In \cite{zhao2019speech}, a 1D CNN LSTM network and a 2D CNN LSTM network were designed to capture local and global features for SER.
Despite achieving great recognition results, when it comes to the speaker-independent SER task that is more analogous to real-world scenarios, their performance tends to exhibit significant degradation. The primary reason for this lies in the fact that individuals exhibit varying rhythms, tones, and styles when speaking. Even when expressing the same emotion, such as happiness in their utterances, the differences can be substantial. However, speaker-dependent SER methods, in their model design and training stages, have not taken this into account.

As a consequence, an increasing number of recent works have shifted their focus towards the speaker-independent SER task.
In contrast, the speaker-independent SER task poses a more intricate challenge, as it not only requires accurate differentiation of diverse speech emotions but also demands the resolution or alleviation of feature distribution variations introduced by different speakers during training and testing phases.
To solve this problem, numerous studies have been advocated. 
For example, 
\cite{chen20183} proposed an attention-based convolutional recurrent neural network  to extract discriminative reprsentations for emotion recognition using Mel-spectrogram with deltas and delta-deltas.
In \cite{gat2022speaker}, the authors presented a speaker normalization method based on the pre-trained HuBERT features for SER.

\subsection{Adversarial Domain Adaptation}
The generalization of SER systems is particularly crucial within the SER domain, because it delineates the systems' viability for deployment in real-world contexts.

Therefore, as one of the promising means to enhance the generalisation performance, adversarial domain adaptation (ADA) approaches \cite{hh2018domain,zhou2019transferable,xiao2020learning,ahn2021cross,latif2022self} have gained much attention. 
Holistically, existing ADA approaches aim to enhance the performance of the SER methods by controlling various speech attributes of the speech signal via techniques such as multi-task learning, adversarial learning, and so on.
For instance,
Xiao \emph{et al} \cite{xiao2020learning} presented a generalized domain adversarial neural network (GDANN) and a class-aligned GDANN to learn the domain-invariant feature representations for robust emotion recognition. 
Latif \emph{et al} \cite{latif2022self} introduced an adversarial dual discriminator (ADDi) network and a self-supervised learning based ADDi network to address the performance degradation of the cross-corpus and cross-language SER. 
Ahn \emph{et al} \cite{ahn2021cross} proposed a few-shot learning and unsupervised domain adaptation based SER model, aiming to acquire emotion similarity by leveraging source samples adapted to the target domain. 

Nonetheless, despite improved performance, they usually neglect the interconnections among various speech attributes.
In contrast, our study endeavors to augment speech emotion modeling through a holistic consideration of the interrelationships among diverse speech attributes.

\subsection{Out-of-distribution Detection}
With continuously deployment of deep learning techniques in practical applications, their reliability has become a growing concern, especially in the safety-critical and security-critical scenarios \cite{berend2020cats}. 

Hence, as one of fundamental tasks for ensuring the reliability of DNN-based systems, out-of-distribution (OoD) detection has attract the attention of numerous researchers.  
Generally speaking, OOD detection aims to enable the model to distinguish between known in-distribution (ID) samples and unseen OOD samples, while not changing the original architecture \cite{lang2023survey,yang2024generalized}. 
Most of them \cite{hendrycksbaseline,lee2018simple,sun2021react,hendrycks22a} will not compromise the original capabilities of the model, except for those attempting to establish better calibrated OOD models early in the training phase \cite{huang2021mos,wei2022mitigating}.
At present, research on OOD detection predominantly concentrates on the fields of CV \cite{hendrycksbaseline,lee2018simple,huang2021mos,sun2021react,hendrycks22a} and NLP \cite{zheng2020out,zeng2021modeling,aroraetal2021types,song2023deeplens}.
On the contrary, there is relatively limited research on OOD detection in the domain of speech processing, with the existing studies primarily centered around the speech recognition task \cite{qiu2021multitask,9739834,li2022improving}.

\section{Methodology}
\label{sec:METHODOLOGY}

This section delineates the detailed configuration of the proposed unified SER framework, MSAC-SERNet. 
Comprising three integral components, namely the input pipeline, the proposed SER model, and the MSAC learning paradigm, Fig. 1 illustrates the overall architecture. 
Further details about each component will be elaborated below.

\subsection{Input Pipeline}
In contrast to prevalent SER methods that typically use MFCCs, we choose to utilize FilterBank features (FBanks) extracted from speech signals as input for the proposed MSAC-SERNet workflow. 
Because we believe that, compared to MFCCs, FBanks possess higher information density and pay greater attention to frequency energy, making them potentially more suitable for SER tasks. To be specific, all speech signals undergo normalization within the range of -1 to 1. Subsequently, we apply the framing operation and a Hamming window to segment each speech signal into frames with a length of 25ms and a shift of 10ms. Finally, we obtain the FBanks for each frame through a 1024-point fast Fourier transform and a mel-scale filter bank analysis.

\subsection{Discriminative Feature Extraction}

Instinctively, the feature extraction component holds a pivotal role in the design of speech emotion modeling, since the quality of the acquired features pose an important impact on the final performance. 
In consequence, we design two shallow and deep feature extraction modules.

Specifically, we first designed three parallel convolution modules, each incorporating diverse receptive field sizes.
This strategic design aims to guarantee the comprehensive utilization of spatial and temporal information inherent in the input Fbanks by the model.
To be explicit, these three convolution modules are composed of parallel convolution layers with kernel sizes of $11\times$1, $3\times$3, and $9\times$1. Following this convolutional operation, one batch normalization (BN) and one rectified linear unit (ReLU) activation layers are applied. 
The resulting features undergo the concatenation operation and an additional convolution layer with a $5\times$5 kernel, followed by a downsampling operation through an average pooling layer.

Subsequently, in order to further augment the representational capacity and discriminability of shallow features, we additionally constructed a deep feature extraction module which consists of multiple convolution blocks.
Its specific architecture is shown in Table. \ref{tab:deep-feature-extraction}.
\begin{table}[htbp]
    \centering
    \caption{Details of the Deep Feature Extraction Module.}
    \label{tab:deep-feature-extraction}
    \renewcommand{\arraystretch}{1.0} 
    \begin{tabular}{>{\centering\arraybackslash}m{15mm}>{\centering\arraybackslash}m{15mm}>{\centering\arraybackslash}m{15mm}>{\centering\arraybackslash}m{15mm}}
        \toprule
        block name         &  kernel\_size     &  channel\_{in}        &  channel\_{out} \\
        \midrule
        \multirow{2}{*}{conv block1}        &  $5\times$5     &    96     &     32   \\
                                            &  $3\times$3     &    32     &     32   \\
        \midrule
        \multirow{2}{*}{conv block2}        &  $3\times$3     &    32     &     64   \\
                                            &  $1\times$1     &    64     &     64   \\
        \midrule
        \multirow{2}{*}{conv block3}        &  $3\times$3     &    64     &     128   \\
                                            &  $1\times$1     &    128     &     128   \\
        \midrule
        \multirow{2}{*}{conv block4}        &  $3\times$3     &    128     &    256    \\
                                            &  $1\times$1     &    256     &    256    \\
        \midrule
        \multirow{2}{*}{conv block5}        &  $3\times$3     &    256     &    256    \\
                           &  $1\times$1     &    256     &    256    \\
        \bottomrule
    \end{tabular}
\end{table}

\subsection{Aggregation Pooling}
After capturing the emotional feature representations, we constructed an aggregation pooling component based on statistical information, so as to eliminate the influence of time dimension features on the final emotion classification. 

\begin{figure}[h]
\centering
	\includegraphics[height=3.8cm,width=!]{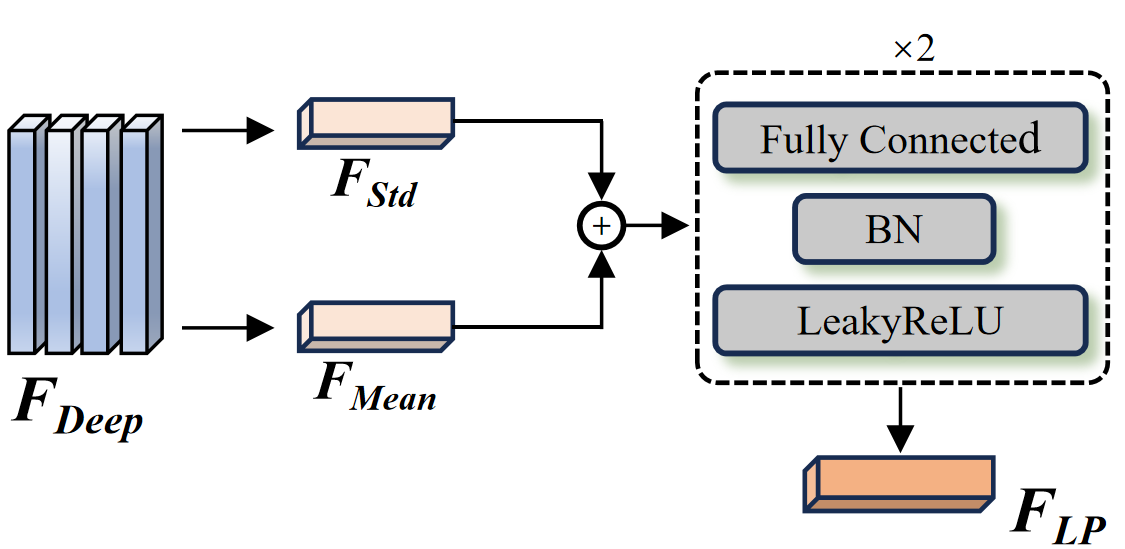}
	\caption{\label{fig:aggpool} Schematics of the proposed aggregation pooling module.}
\end{figure}

To be concrete, considering the importance of the variation and energy of speech signals for emotion recognition, we first extract and concatenate the mean and standard deviation of the temporal dimension within the emotional features .
Then, a linear projection module which consists of two blocks is built to better leverage the above captured features. As illustrated in Fig. \ref{fig:aggpool}, each block contains one fully connected layer, one BN layer, and one LeakyReLU layer.

\subsection{Loss Function}
Normally, SER methods employ cross-entropy loss for model training, whereas we contend that this approach is prone to inducing emotional confusion. 

To mitigate this concern, we choose to adopt the additive margin softmax (AM-Softmax) loss to optimize our MSAC-SERNet, instead of employing conventional regularization methods like label smoothing. 
By doing so, the proposed SER model s able to enlarge the distance between different classes and minimize the distance within the same category.
\begin{equation}
    \begin{split}
        \cos(\theta_{j, i}) = \frac{{x}_{i}^{T}w_j}
        {\left\Vert x_i \right\Vert \left\Vert w_j \right\Vert}
    \end{split}
\end{equation}
\begin{equation}
    \begin{split}
        L_E = -\frac{1}{N}\sum_{i=1}^N \log
        \frac{e^ {s(\cos(\theta_{y_i, i}) - m)}}
        {e^{s(\cos(\theta_{y_i, i}) - m)} + \sum_{j\neq{y_i}}e^{s(\cos(\theta_{j, i}))}} 
    \end{split}
\end{equation}
\noindent 
where $L_{E}$ represents the final loss of the emotion attribute, $x_i$ and $y_i$ indicate the feature vector and label of the $i$th sample, $w_j$ is the feature vector of class $j$, $\theta_{j, i}$ denotes the angle between $x_i$ and $w_j$, N denotes the batch size, s is the scaling factor, m is the additive margin. In our case, m and s are set as 0.2 and 30. 

\subsection{Multiple Speech Attribute Control Method}

\begin{figure}[h]
\centering
	\includegraphics[height=8.8cm,width=!]{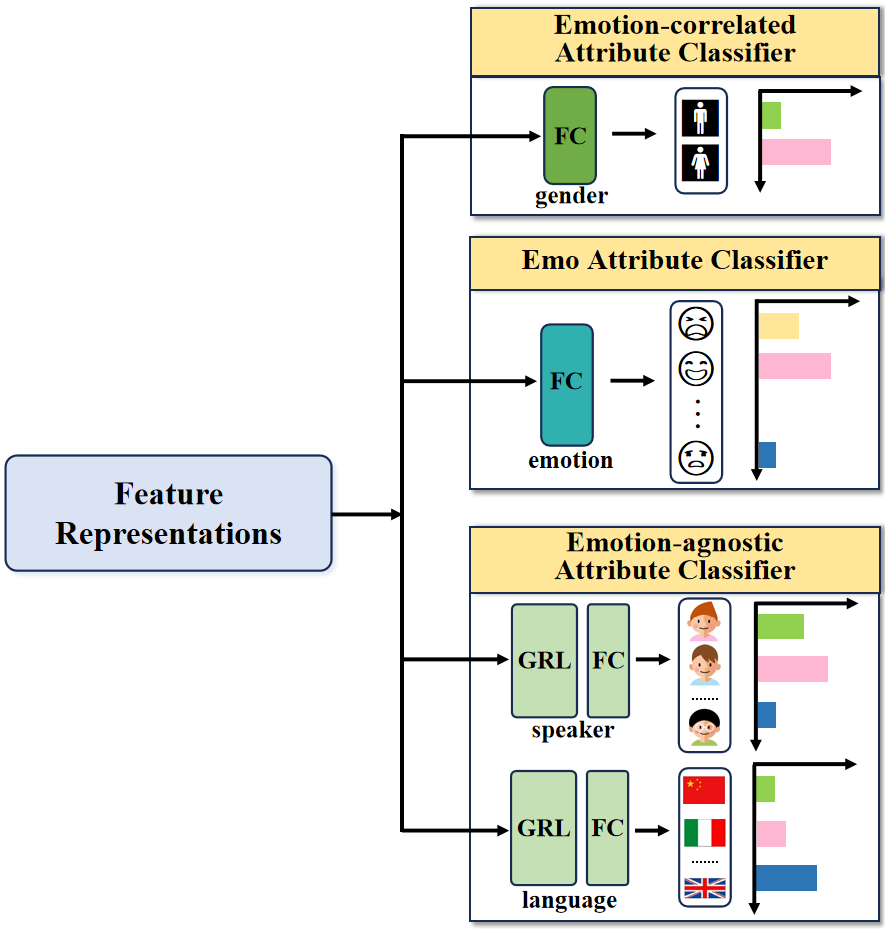}
	\caption{\label{fig:msac} Overview of the proposed MSAC learning paradigm. FC denotes the fully connected layer, GRL represents the gradient reversal layer.}
\end{figure}

Existing studies \cite{wang2023neural,kharitonov2023speak,yao2023promptvc} have demonstrated the intricate challenge of decoupling speech signals, highlighting the inevitable overlap among different speech attributes. 
In this context, numerous researchers \cite{hh2018domain, gat2022speaker, latif2022self} have endeavored to enhance SER performance through the incorporation of attribute-based ADA methods. 
Nevertheless, we contend that their performance still falls short of optimal, as they predominantly utilize diverse speech attributes without adequately considering their interrelationships.
To solve this issue, a novel and effective MSAC method that explicitly models and controls different speech attributes is proposed, as shown in Fig. \ref{fig:msac}. 

Unlike previous works, our proposed MSAC models speech emotion attribute based on correlations across additional speech attributes, aiming to simultaneously capture fine-grained emotion-related representations while mitigating the negative impact of emotion-agnostic features.
More precisely, our proposed MSAC method first classifies different speech attributes into \textbf{emotion}, \textbf{emotion-agnostic} and \textbf{emotion-correlated} attributes.
The speaker and language speech attributes are regarded as emotion-agnostic speech attributes, because they share fewer acoustic properties with the emotion attribute. 
In contrast, the gender attribute is an emotion-correlated speech attribute, as male and female speech signals usually exhibit substantial differences in energy, pitch, and other acoustic characteristics which are of great importance for SER.
Then, for emotions and emotion-associated attributes, we employ a multi-task learning strategy to preserve valuable information for emotion identification, guided by the AM-Softmax loss. 
Meanwhile, regarding emotion-agnostic speech attributes, we not only utilize the multi-task learning strategy but also employ gradient reversal \cite{hh2018domain} which transmits data during forward propagation and reverses the sign of the gradient during backward propagation, thereby mitigating their influence on the emotion attribute.
Moreover, it is essential to point out that all of these speech attributes should be optimized together since they are mutually influenced.

Therefore, the final classification loss can be formulated as follows:
\begin{equation}
    \begin{split}
        L_{E_{ag}} = \sum_{i=1}^N \alpha_{i}L_{a_{i}}
    \end{split}
\end{equation}
\begin{equation}
    \begin{split}
        L_{E_{co}} = \sum_{j=1}^M \alpha_{j}L_{a_{j}}
    \end{split}
\end{equation}
\begin{equation}
    \begin{split}
        L_{total} = (1-\sum_{j=1}^N\alpha_{i}-\sum_{j=1}^M\alpha_{j}) L_{E} + L_{E_{ag}} + L_{E_{co}}
    \end{split}
\end{equation}
where $L_{E_{ag}}$, $L_{E_{co}}$, and $L_{E}$ are AM-Softmax classification loss of emotion-agnostic, emotion-correlated, and emotion attributes, $a_i$ and $\alpha_i$ are emotion-agnostic attribute and its weight coefficient, $a_j$ and $\alpha_j$ are emotion-correlated attribute and its weight coefficient, $N$ and $M$ represent the number of used emotion-agnostic and emotion-correlated attributes. 
Regarding the values for these hyper-parameters, please refer to the experiment section.

As a consequence, the proposed SER model using MSAC is capable of extracting more discriminative emotional feature representations, thus improving its overall performance.

\section{Experimental Databases And Setup}
\label{sec:EXPERIMENTS}

In this section, the used experimental datasets for the single-corpus and cross-corpus SER tasks are first introduced. 
Then, the detailed configurations of all experiments are given. 
Last, we describe the overall experimental results and analysis.

\subsection{Dtatbases}
 
To verify the effectiveness of our proposed unified SER workflow, we totally use seven public speech emotion corpora and perform extensive experiments on both single-corpus and cross-corpus SER tasks. Details of these datasets are as follows:

\begin{enumerate}
    \item IEMOCAP \cite{IEMOCAP}: IEMOCAP is a key resource in the realm of SER, consists of 12 hours of multimodal data featuring 10 performers. Across five sessions, both male and female actors participate in scripted and improvised scenes, providing transcriptions, speech waveforms, and visual frames for comprehensive analysis. IEMOCAP offers diverse emotional expressions, including angry, happy, sad, surprise, fear, disgust, neutral, frustration, and excitement. All speech signals are sampled at a rate of 16 kHz with 16-bit resolution.
    \item EMO-DB \cite{EMODB}: EMO-DB is a significant German emotional database which comprises 10 speakers delivering 535 utterances in the German language, totaling 25 minutes of audio. EMO-DB is characterized by 7 emotions which are angry, boredom, disgust, fear, happy, neutral, sad.
    \item EMOVO \cite{EMOVO}: EMOVO, an Italian emotional dataset, features 6 speakers (3 males and 3 females), presenting 588 utterances at a 48 kHz sampling rate. This corpus encompasses 7 emotions, which are angry, disgust, joy, fear, sad, neutral, and surprise.
    \item CASIA \cite{CASIA}: CASIA, a Chinese emotion corpus, showcases recordings from four professional speakers (2 male and 2 female), portraying six emotions: angry, happy, fear, sad, surprise, and neutral. The dataset comprises 1,200 public sentences, sampled at a rate of 16 kHz.
    \item RAVDESS \cite{RAVDESS}: RAVDESS, the Ryerson Audio Visual Database of Emotional Speech and Song, encompasses 24 speakers delivering 2542 English utterances. The corpus has total 2 hours and 47 minutes of audio, with 8 emotions: happy, sad, angry, fear, surprise, disgust, calm, and neutral.
    \item TESS \cite{TESS}: TESS, the Toronto Emotional Speech Set, comprises two actors (one male and one female), presenting 2800 English emotional utterances. This dataset, amounting to 1 hour and 36 minutes of audio, encompasses seven emotions: angry, disgust, joy, fear, sad, neutral, and happy.     
\end{enumerate}

To be explicit, in single-corpus SER, we first select the most challenging IEMOCAP corpus.
To make fair comparisons, we likewise select four classes (happy \!+\! excited, angry, sad, and neutral) to evaluate the recognition performance of the proposed SER workflow. 
The remaining three classes (frustrated, fear, and surprised) are chosen to test the reliability performance of the proposed MSAC-SERNet.
For a comprehensive comparison, we also choose EMO-DB to evaluate MSAC-SERNet's recognition performance due to its widespread utilization.

As for the cross-corpus SER task, five classes (happy, angry, sad, fear, and neural) of five gender-balanced corpora (EMO-DB, EMOVO, IEMOCAP, CASIA, and RAVDESS) are employed to verify the recognition performance of the proposed SER method, while their remaining classes are utilized to examine the reliability performance of the proposed MSAC-SERNet.
In terms of the generalization performance, the same five emotion classes of one unseen corpus TESS are chosen.

% Details are illustrated in Table \ref{tab:data-details}.

% \begin{table}[htbp]
%     \centering
%     \caption{Details of the experimental dataset for cross-corpus SER.}
%     \label{tab:data-details}
%     \renewcommand{\arraystretch}{1.0} 
%     \begin{tabular}{ccccccc}
%         \toprule
%         Datasets          & Angry   & Happy  & Neutral  & Sad   & Fear  & Total \\
%         \midrule
%         EMO-DB             &  1103   &  83      \\
%         EMOVO             &  1636   &  126     \\
%         IEMOCAP           &  1708   &  111     \\
%         CASIA               &  1084   &  99      \\
%         RAVDESS               &  1084   &  99      \\
%         TESS               &  1084   &  99      \\
%         \midrule
%         Total             &  5531   &  419      \\
%         \bottomrule
%     \end{tabular}
% \end{table}

\subsection{Implementation Details}
In all experiments, AdamW \cite{loshchilov2019decoupled} is adopted to optimize all models with an initial learning rate of 0.001 and a batch size of 64, and they are all trained for 100 epochs. 
Besides, SpecAugment \cite{specaugment} is applied to input Fbanks with a frequency mask width of 2 as well as a time mask width of 30 frames. 

Concerning the best configurations of weight coefficients, 
we set $\alpha_{E}$=0.5, $\alpha_{S}$=0.3, $\alpha_{G}$=0.2 for IEMOCAP, 
$\alpha_{E}$=0.85, $\alpha_{S}$=0.1, $\alpha_{G}$=0.05 for EMODB, 
and $\alpha_{E}$=0.5, $\alpha_{S}$=0.15, $\alpha_{G}$=0.2, $\alpha_{L}$=0.15 for cross-corpus SER.

For a fair comparison with SOTA SER methods, we adopt the standard 10-fold speaker-independent cross-validation in the single-corpus SER task, which is the same as existing studies \cite{zhong2020lightweight,peng2021efficient,hou2021multi,fan2022isnet}.
As for the cross-corpus SER task, the holdout speaker-independent cross-validation is employed. 
Concretely, two randomly selected speakers from each experimental datasets of cross-corpus SER are respectively employed as the valid and test sets, while the remaining data of each datasets is utilized as the training set.

\subsection{Evaluation Metrics}

Different metrics are used to validate the recognition, generalization, and reliability performance of the proposed SER workflow. 

To be precise, the weighted average recall (WAR) and unweighted average recall (UAR) are employed to evaluate the recognition and generalization performance.
% 加个公式
In terms of the reliability performance, the false positive rate at 95\% true positive rate (FPR95) and Area Under the Receiver Operating Characteristic curve (AUROC) are adopted.
% 加个公式

\section{Experimental Results And Discussion}
\subsection{Comparison With Existing Works}

\subsubsection{Single-corpus SER Task}
To demonstrate the effectiveness of our proposed unified SER workflow MSAC-SERNet, we first compare its recognition performance with recent SOTA SER methods on the IEMOCAP and EMO-DB datasets. The results are shown in Table \ref{tab:IEMOCAP-results} and Table \ref{tab:EMODB-results}.

\begin{table}[htbp]
    \centering
    \caption{Recognition comparison of SOTA SER methods on IEMOCAP. The ‘-’ represents the lack of this measure, and the best performance is bolded.}
    \label{tab:IEMOCAP-results}
    \renewcommand{\arraystretch}{1.0} 
    \begin{tabular}{>{\centering\arraybackslash}m{30mm}>{\centering\arraybackslash}m{10mm}>{\centering\arraybackslash}m{9mm}>{\centering\arraybackslash}m{9mm}}
        \toprule
        Algorithms         &  Year   &   UAR   &  WAR   \\
        \midrule
        CNN+Bi-GRU \cite{zhong2020lightweight}   & 2020 &  71.72  &  70.39  \\
        \midrule
        SPU-MSCNN \cite{peng2021efficient}       & 2021 &  68.40  &  66.60  \\
        \midrule
        Light-SERNet \cite{9746679}              & 2022 &  70.76  &  70.23  \\
        \midrule
        ISNet \cite{fan2022isnet}                &  2022 &  70.43  &  65.02  \\
        \midrule
        CMRN \cite{hou2021multi}                 &  2022 &  66.64  &  -  \\
        \midrule
        SS-AAE \cite{latif2020multi}               &  2022 &  66.70  &  -  \\
        \midrule
        Dual-TBNet \cite{liu2023dual}            &  2023 &  64.80  &  -  \\
        \midrule
        Ours          &  2023    &  \textbf{71.76}  &  \textbf{72.97}  \\
        \bottomrule
    \end{tabular}
\end{table}

\begin{table}[htbp]
    \centering
    \caption{Recognition comparison of SOTA SER methods on EMO-DB. The ‘-’ denotes the lack of this measure. The best performance is bolded and the second best is underlined.}
    \label{tab:EMODB-results}
    \renewcommand{\arraystretch}{1.0} 
    \begin{tabular}{>{\centering\arraybackslash}m{30mm}>{\centering\arraybackslash}m{10mm}>{\centering\arraybackslash}m{9mm}>{\centering\arraybackslash}m{9mm}}
        \toprule
        Algorithms         &  Year   &   UAR   &  WAR   \\
        \midrule
        SGMM-HMM \cite{mao2019revisiting}   & 2019 &  90.48  &  88.25  \\
        \midrule
        CNN-FA \cite{peng2021efficient}       & 2021 &  83.30  &  82.10  \\
        \midrule
        DIFL  \cite{lu2022domain}              & 2022 &  89.72  &  88.49  \\
        \midrule
        CMRN \cite{hou2021multi}               &  2022 &  \textbf{92.51}  &  -  \\
        \midrule
        Dual-TBNet \cite{liu2023dual}          &  2023 &  84.10  &  -  \\
        \midrule
        TFPE-SVM  \cite{mishra2023chirplet}     & 2023 &  85.60  &  -  \\
        \midrule
        RH-emo+Quat \cite{guizzo2023learning}    &  2023 &  73.00  &  65.64  \\
        \midrule
        CNN-CasA-Tri \cite{liu2023discriminative}  &  2023 &  91.58  &  88.76  \\
        \midrule
        Ours          &  2023  &  \underline{92.11}  &  \textbf{93.21}  \\
        \bottomrule
    \end{tabular}
\end{table}

From Table \ref{tab:IEMOCAP-results} and Table \ref{tab:EMODB-results}, it can be easily seen that our proposed MSAC-SERNet achieves the best recognition performance among other SOTA methods.
To be exact, MSAC-SERNet obtains the best WAR of 72.97\% and UAR of 71.76\% on IEMOCAP, and acquire the best WAR of 93.21\% and secondary best UAR of 92.11\% on EMO-DB, respectively. 
These achievements are attained through the fine-grained modeling of the emotion attribute present in speech signals.
On one hand, our proposed MSAC learning paradigm takes into account the intricate interplay among diverse speech attributes, which enables the extraction of feature representations that encompass emotional information from speech signals to the fullest extent possible.
On the other hand, the designed SER model not only captures discriminative features from speech signals, but extends the inter-class distance between different emotion categories while decreasing the  intra-class distance within the same emotion category.

\subsubsection{Cross-corpus SER Task}

For a fair comparison with other approaches, we present existing learning paradigms that share a similar conceptual framework as outlined in this paper, along with their respective performance metrics in Table \ref{tab:ccser-results}.

\begin{table}[htbp]
    \centering
    \caption{Recognition comparison of SOTA SER methods on Cross-corpus SER. Base represents the proposed SER model.}
    \label{tab:ccser-results}
    \renewcommand{\arraystretch}{1.0} 
    \begin{tabular}{>{\centering\arraybackslash}m{24mm}>{\centering\arraybackslash}m{6mm}>{\centering\arraybackslash}m{5mm}>{\centering\arraybackslash}m{5mm}>{\centering\arraybackslash}m{20mm}}
        \toprule
        \multirow{2}{*}{Models} & \multirow{2}{*}{Year} & \multicolumn{2}{c}{ID Recognition} & \multicolumn{1}{c}{OOD Generalization} \\
        \cmidrule(lr){3-4} \cmidrule(lr){5-5} 
            &    & WAR  &  UAR  &  WAR/UAR    \\
        \midrule
        Base               & 2023        & 44.64       & 44.42  &  54.25   \\
        \midrule
        Base+GMT\cite{li2019improved}         & 2019     & 48.95   & 48.27   &  52.05    \\
        Base+TAP\cite{gat2022speaker}         & 2022     & 49.22       & 46.98  & 57.55    \\
        Base+AGR\cite{parry2022speech}         & 2022     & 49.68       & 48.83  & 47.15    \\
        Base+AMTL\cite{parry2022speech}        & 2022     & 52.98       & 52.39  & 54.75    \\
        \midrule
        Ours (Base+MSAC)   & 2023  & \textbf{55.18}  & \textbf{53.67}  & \textbf{70.00}   \\
        \bottomrule
    \end{tabular}
\end{table}

From Table \ref{tab:ccser-results}, it is evident that our proposed MSAC consistently achieves the best ID recognition and OOD generalization results. 
\begin{itemize}
    \item Concerning ID recognition performance, our base SER model achieves a WAR of 44.64\% and a UAR of 44.42\%.     
    When incorporating recent learning paradigms \cite{li2019improved, gat2022speaker, parry2022speech}, their WAR and UAR are all improved. Among them, AMTL demonstrated superior performance, with a 8.34\% increase in WAR and a 7.97\% increase in UAR. 
    In comparison, our proposed MSAC achieves the best ID recognition performance, with absolute improvements of 10.54\% WAR and 9.25\% UAR.
    \item 
    In terms of the OOD generalization performance, the proposed SER base model attains a WAR/UAR of 54.25\%.
    When using other speech attributes as auxiliary task, GMT and AGR exhibit lower performance compared to the proposed base model, while TAP and AMTL demonstrate slight performance enhancements.
\end{itemize}

\vspace{-3mm}
\subsection{Ablation Study}
To comprehensively assess and analyze the validity of our proposed MSAC-SERNet and its components, more detailed ablation study is performed.

\subsubsection{Single-corpus SER Task}
We first compare the recognition performance of our MSAC-SERNet and its variants on IEMOCAP and EMODB within single-corpus SER, as presented in Table \ref{tab:ablation-recognition-IEMOCAP} and Table \ref{tab:ablation-recognition-EMODB}.

\begin{table}[htbp]
    \centering
    \caption{\label{tab:ablation-recognition-IEMOCAP} Recognition Comparison of our proposed MSAC-SERNet and its variants on IEMOCAP.}
    \renewcommand{\arraystretch}{1.0} 
    \begin{tabular}{>{\centering\arraybackslash}m{30mm}>{\centering\arraybackslash}m{15mm}>{\centering\arraybackslash}m{15mm}}
        \toprule
        \multirow{2}{*}{Models} & \multicolumn{2}{c}{ID Recognition} \\
        \cmidrule(lr){2-3} 
            & WAR  &  UAR  \\
        \midrule 
        Base                            & 68.40       & 68.42   \\
        Base+$S{\!}_{G{\!}R}$           & 69.23       & 68.33   \\
        Base+$G{\!}_{M{\!}L}$           & 69.02       & 68.10   \\
        Ours (Base+MSAC)                & \textbf{72.97}  & \textbf{71.76}  \\
        \bottomrule
    \end{tabular}
\end{table}

\begin{table}[htbp]
    \centering
    \caption{\label{tab:ablation-recognition-EMODB} Recognition Comparison of our proposed MSAC-SERNet and its variants on EMODB.}
    \renewcommand{\arraystretch}{1.0} 
    \begin{tabular}{>{\centering\arraybackslash}m{30mm}>{\centering\arraybackslash}m{15mm}>{\centering\arraybackslash}m{15mm}}
        \toprule
        \multirow{2}{*}{Models} & \multicolumn{2}{c}{ID Recognition} \\
        \cmidrule(lr){2-3} 
            & WAR  &  UAR  \\
        \midrule
        Base                      & 85.71       & 80.48   \\
        Base+$S{\!}_{G{\!}R}$     & 88.57       & 83.33   \\
        Base+$G{\!}_{M{\!}L}$     & 87.76       & 87.25   \\
        Ours (Base+MSAC)          & \textbf{92.11}  & \textbf{93.21}  \\
        
        \bottomrule
    \end{tabular}
\end{table}

We can easily observe that with the increase of additional speech control, the ID recognition performance of the SER mdoel is gradually enhanced. The rationale behind this trend stems from the reasonable and stringent constraints imposed by learning paradigms, allowing the model to incorporate more useful information from diverse speech attributes. This approach, to some extent, facilitates the effective acquisition of speech emotion information, leading to an overall enhancement in the model's performance.
Besides, as evident from the tables above, the results clearly demonstrate the superior performance of our proposed MSAC over existing learning 
\begin{figure*}[h]
\centering
	\includegraphics[height=5cm,width=!]{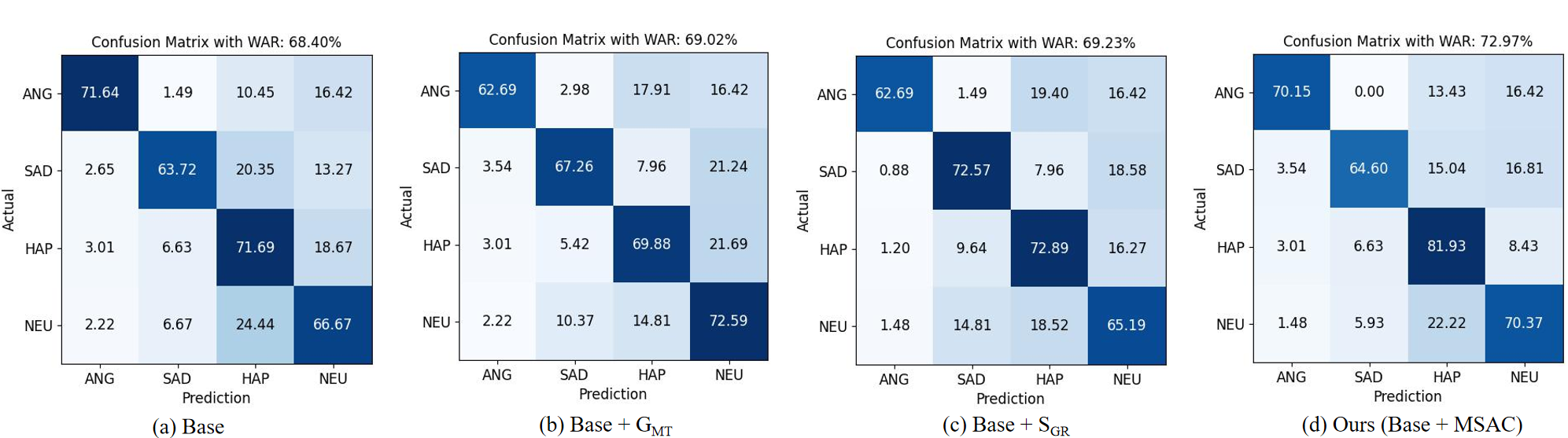}
	\caption{\label{fig:iemocap} Confusion matrices for the variants of the proposed MSAC-SERNet on IEMOCAP. The diagonal numbers represent the recall rate for each emotion.}
\end{figure*}
paradigm such as \cite{parry2022speech}. This superiority is attributed to the fact that our MSAC is modeled based on the relationships among diverse speech attributes, representing a more rational approach that enables the capture of finer-grained emotional feature representation.

To scrutinize the recognition performance of each category, we further present the confusion matrices for MSAC-SERNet and its variants on IEMOCAP in Fig. \ref{fig:iemocap}, where ANG, SAD, HAP, NEU represent angry, sad, happy, and neutral.
It illustrates that our proposed MSAC-SERNet gains a comparable or higher recognition accuracy for each emotion than its variants, which validates the effectiveness of the proposed SER model architecture and MSAC learning paradigm.

\subsubsection{Cross-corpus SER Task}
Moreover, We also test the recognition and generalization performance of our proposed MSAC-SERNet and its variants in the cross-corpus SER task.

\begin{table}[htbp]
    \centering
    \caption{\label{tab:ablation-recognition-cross} Recognition and generalization results of our proposed MSAC-SERNet and its variants in cross-corpus SER.}
    \renewcommand{\arraystretch}{1.0} 
    \begin{tabular}{>{\centering\arraybackslash}m{35mm}>{\centering\arraybackslash}m{6mm}>{\centering\arraybackslash}m{6mm}>{\centering\arraybackslash}m{20mm}}
        \toprule
        \multirow{2}{*}{Models} & \multicolumn{2}{c}{ID Recognition} & \multicolumn{1}{c}{OOD Generalization} \\
        \cmidrule(lr){2-3} \cmidrule(lr){4-4} 
             & WAR  &  UAR  &  WAR/UAR    \\
        \midrule
        Base                                      & 44.64       & 44.42   &  54.25   \\
        \midrule
        Base+$S{\!}_{G{\!}R}$                     & 49.22       & 46.98   &  57.55    \\
        Base+$G{\!}_{M{\!}L}$                     & 48.95       & 48.27   &  52.05    \\
        Base+$L{\!}_{G{\!}R}$                     & 47.20       & 46.47   &  56.10    \\
        Base+$S{\!}_{G{\!}R}$+$G{\!}_{M{\!}L}$    & 51.24       & 52.87   &  58.35    \\
        Base+$S{\!}_{G{\!}R}$+$L{\!}_{G{\!}R}$    & 49.95       & 48.74   &  56.05    \\
        Base+$L{\!}_{G{\!}R}$+$G{\!}_{M{\!}L}$    & 51.24       & 50.74   &  53.65    \\
        \midrule
        Ours (Base+MSAC)                          & \textbf{55.18}  & \textbf{53.67}  & \textbf{70.00}   \\
        \bottomrule
    \end{tabular}
\end{table}

As illustrated in Table \ref{tab:ablation-recognition-cross}, it can be ascertained that the proposed MSAC-SERNet also consistently achieves the best performance in terms of ID recognition and OOD generalization performance, which further validates the effectiveness of the proposed unified SER workflow.

Concretely, in terms of the ID recognition and OOD generalization performance, our proposed SER model achieves a ID WAR of 44.64\%, a ID UAR of 44.42\%, and a OOD WAR/UAR of 54.25\%, respectively.
Then, when applying the commonly used ADA method which could enable the model to be less affected by the emotion-agnostic speaker attribute, the WAR and UAR of the model's ID recognition performance are improved by 4.58\% and 2.56\%, the WAR/UAR of the model's OOD generalization performance is improved by 3.3\%.
Afterward, when adding $G{\!}_{M{\!}L}$ stragegy on top of SGR, the model is able to reduce the negative impact of the speaker attribute while retaining the useful emotion-correlated gender attribute information for SER. 
As a result, the WAR and UAR of ID recognition performance are further enhanced by 2.02\% and 5.89\%, the WAR/UAR of OOD generalization performance is further improved by 0.8\% as well.
In particular, when using our proposed MSAC method which simultaneously controls all known emotion-agnostic and emotion-correlated speech attributes, the proposed SER model achieves the best ID recognition and OOD generalization performance which outperforms the baseline by 8.61\%, 9.29\% and 7.35\%, respectively.

\subsection{Reliability Comparison And Analysis}

In recent years, as AI models continue to showcase substantial potential across diverse practical domains, their reliability has garnered significant attention. 
In this context, we initiate a preliminary analysis of the reliability of SER models when facing semantic data shifts.
To be specific, we use four SOTA OOD detection methods and propose one simple OOD detection method rODIN which is a variant of ODIN, so as to verify the reliability performance of our MSAC-SERNet.

\subsubsection{Single-corpus SER Task}

First, we compare the reliability of the proposed MSAC-SERNet and its variants in the single-corpus SER task, whose corresponding results are delineated in Table \ref{tab:reliability-results-sser}.

\begin{table}[htbp]
    \centering
    \caption{\label{tab:reliability-results-sser} Reliability Comparison of our proposed MSAC-SERNet and its variants in Single-corpus SER. The optimal outcomes for each OOD detection method are emphasized in bold, with the overall best results across all methods highlighted in red.}
    \renewcommand{\arraystretch}{0.9}
    \begin{tabular}{>{\centering\arraybackslash}m{18mm}>{\centering\arraybackslash}m{24mm}>{\centering\arraybackslash}m{13mm}>{\centering\arraybackslash}m{15mm}}
        \toprule
         OOD Detector    &   Model   &   FPR95 $(\downarrow)$    &   AUROC $(\uparrow)$  \\
        \midrule
        \multirow{4}{*}{rODIN}         & Base                       &  95.74  &  59.95   \\
                                       & Base+$S{\!}_{G{\!}R}$      &  81.36  &  63.83   \\
                                       & Base+$G{\!}_{M{\!}L}$      &  93.49  &  62.91   \\
                                       & Ours (Base+MSAC)                      &  \colorbox{lightred}{\textbf{68.29}}  &  \colorbox{lightred}{\textbf{73.23}}   \\
        \midrule[\heavyrulewidth]
        \multirow{4}{*}{MaxLogit}      & Base                       &  95.89  &  59.73   \\
                                       & Base+$S{\!}_{G{\!}R}$      &  81.66  &  63.77   \\
                                       & Base+$G{\!}_{M{\!}L}$      &  93.49  &  62.81   \\
                                       & Ours (Base+MSAC)                       &  \textbf{68.49}  &  \textbf{72.95}   \\
        \midrule[\heavyrulewidth]
        \multirow{4}{*}{ODIN}          & Base                       &  96.04  &  59.89   \\
                                       & Base+$S{\!}_{G{\!}R}$      &  81.46  &  63.93   \\
                                       & Base+$G{\!}_{M{\!}L}$      &  92.74  &  62.77   \\
                                       & Ours (Base+MSAC)                       &  \textbf{69.09}  &  \textbf{72.61}   \\
        \midrule[\heavyrulewidth]
        \multirow{4}{*}{ReACT}         & Base                       &  92.23  &  65.15   \\
                                       & Base+$S{\!}_{G{\!}R}$      &  85.47  &  63.64   \\
                                       & Base+$G{\!}_{M{\!}L}$      &  97.34  &  49.59   \\
                                       & Ours (Base+MSAC)                       &  \textbf{78.66}  &  \textbf{65.45}   \\
        \midrule[\heavyrulewidth]
        \multirow{4}{*}{Ma\_Dis}       & Base                       &  100.0  &  41.18   \\
                                       & Base+$S{\!}_{G{\!}R}$      &  98.36  &  42.78   \\
                                       & Base+$G{\!}_{M{\!}L}$      &  97.81  &  37.46   \\
                                       & Ours (Base+MSAC)                       &  \textbf{94.80}  &  \textbf{48.60}   \\
        \bottomrule
    \end{tabular}
\end{table}

According to the reported results, it is apparent that as additional speech attributes are gradually controlled, the reliability performance of the proposed base model continuously improves, with the exception of the $Base+G{\!}_{M{\!}L}$ for the ReACT method. 
In general, the $S{\!}_{G{\!}R}$ performs better than the $G{\!}_{M{\!}L}$ learning paradigm, and our proposed MSAC achieves the best results.
Take the commonly used MaxLogit OOD method as an example. 
Our proposed base model attains FPR95 of 95.74\% and AUROC of 59.95\% which hardly identifies any OOD samples.
When exerting control over additional speech attributes, namely speaker and gender, their reliability performance has been gradually enhanced.
\begin{figure*}
\centering
	\includegraphics[height=13.5cm,width=!]{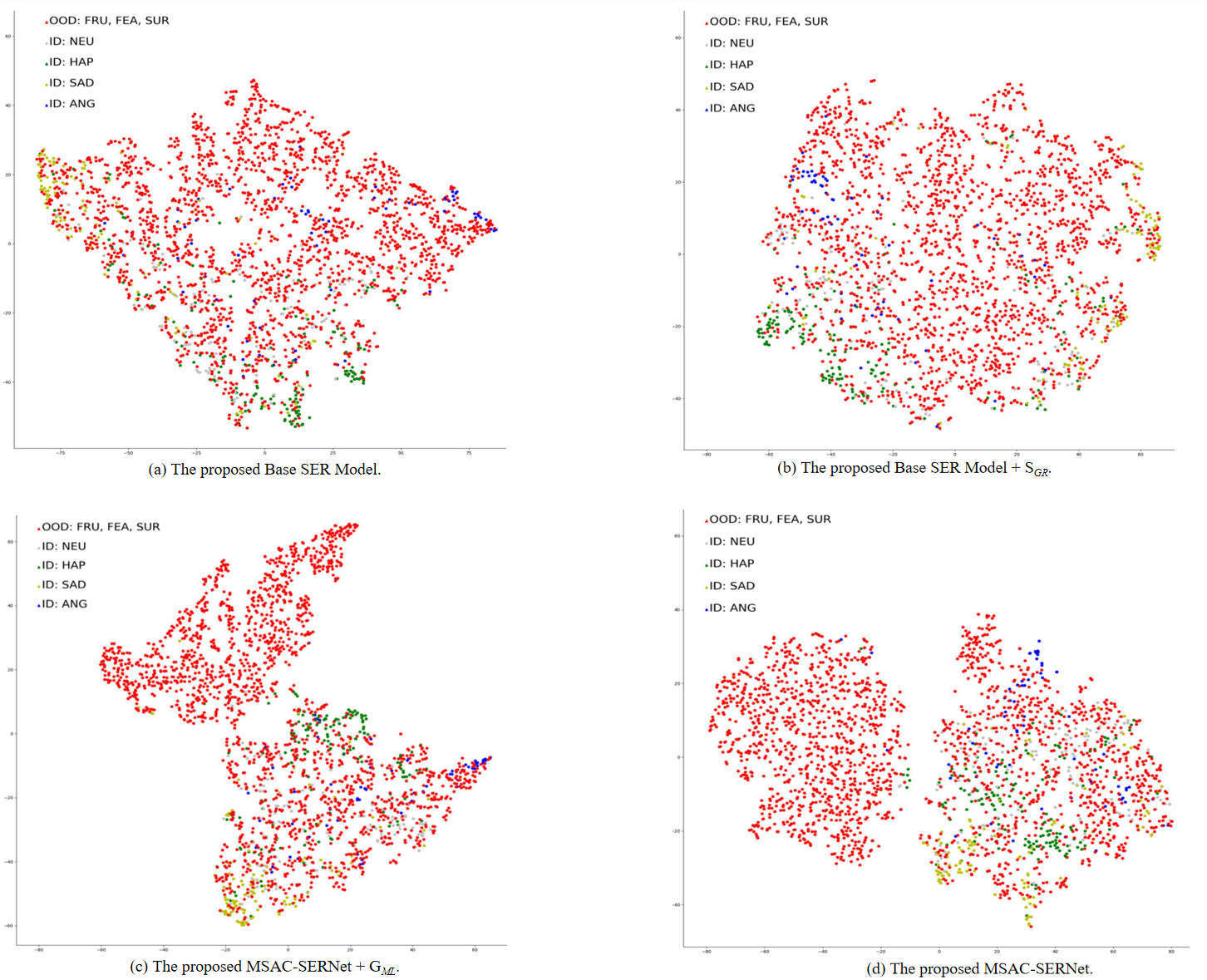}
	\caption{\label{fig:tsne_iemocap_ood} The t-SNE visualization of data distribution on IEMOCAP, where FRU, FEA, SUR represent the emotions of frustration, fear, and surprise.}
 % \caption{\label{fig:tsne_iemocap_ood} The t-SNE visualization of data distribution on IEMOCAP. FRU, FEA, SUR represent the remaining three emotion classes, i.e., frustration, fear, and surprise of the IEMOCAP corpus.}
\end{figure*}
In particular, when utilizing our proposed MSAC learning paradigm, the proposed method realizes a remarkable 27.40\% FPR95 reduction and a significant AUROC upgrade of 13.22\% as opposed to the baseline, reaching a preliminary usable level for OOD detection.
And it also indicates that by incorporating emotion-associated speech attributes and attenuating emotion-agnostic speech information, the proposed MSAC-SERNet can extract more robust emotional features, thereby enhancing the model's OOD detection capabilities when facing semantic shifts.

Additionally, we observe an intriguing phenomenon where commonly used techniques in CV and NLP for enhancing OOD detection performance do not necessarily apply to the SER task. 
Concretely, the ODIN method, which introduces a positive perturbation to the input signal to some extent enlarging the confidence distribution of ID and OOD samples, exhibits undesirable effects in SER. 
Contrarily, our experiments show that our proposed rODIN method which introduces a reverse perturbation to the input signal, akin to generating adversarial examples \cite{goodfellow2015explaining}, can slightly enhance the model's OOD detection capabilities in the domain of SER.

Furthermore, to conduct an in-depth analysis of reliability performance, we employ the unsupervised t-SNE technique to visualize the features' distribution of ID and OOD data from the IEMOCAP corpus. In particular, we perform a comparative analysis using our proposed SER base model, MSAC-SERNet and its variants, as depicted in Fig. \ref{fig:tsne_iemocap_ood}.
From the figure, it is evident that our proposed MSAC-SERNet exhibits a more pronounced disparity in the distribution of ID and OOD data compared to its baseline model and variants. This is attributed to the finer control over various speech attributes incorporated into the MSAC-SERNet, enabling the model to capture more nuanced emotional feature representations. Consequently, this enhancement contributes to improved reliability performance when facing semantic data shifts.

\subsubsection{Cross-corpus SER Task}

Subsequently, the reliability comparison of the proposed MSAC-SERNet and its variants in the cross-corpus SER task is performed, and the results are illustrated in Table \ref{tab:reliability-results-cser}.

\begin{table}[htbp]
    \centering
    \caption{\label{tab:reliability-results-cser} Reliability Comparison of our proposed MSAC-SERNet and its variants in Cross-corpus SER. The optimal outcomes for each OOD detection method are emphasized in bold, with the overall best results across all methods highlighted in red.}
    \renewcommand{\arraystretch}{1.0}
    \begin{tabular}{>{\centering\arraybackslash}m{17mm}>{\centering\arraybackslash}m{24mm}>{\centering\arraybackslash}m{13mm}>{\centering\arraybackslash}m{15mm}}
        \toprule
         OOD Detector    &   Model   &   FPR95 $(\downarrow)$    &   AUROC $(\uparrow)$ \\
        \midrule
        \multirow{8}{*}{rODIN}         & Base                       &  95.34  &  59.10   \\
                                       & Base+$S{\!}_{G{\!}R}$      &  94.48  &  58.95   \\
                                       & Base+$G{\!}_{M{\!}L}$      &  95.37  &  58.78   \\
                                       & Base+$L{\!}_{G{\!}R}$      &  89.23  &  61.26   \\
                                       & Base+$G{\!}_{M{\!}L}$+$S{\!}_{G{\!}R}$      &  91.44  &  55.71   \\
                                       & Base+$G{\!}_{M{\!}L}$+$L{\!}_{G{\!}R}$      &  94.42  &  55.14   \\
                                       & Base+$L{\!}_{G{\!}R}$+$S{\!}_{G{\!}R}$      &  90.82  &  56.43   \\
                                       & Ours (Base+MSAC)                       &  \colorbox{lightred}{\textbf{85.27}}  &  \colorbox{lightred}{\textbf{63.12}}  \\
        \midrule[\heavyrulewidth]
        \multirow{8}{*}{MaxLogit}      & Base                       &  95.28  &  59.08   \\
                                       & Base+$S{\!}_{G{\!}R}$      &  94.25  &  59.00   \\
                                       & Base+$G{\!}_{M{\!}L}$      &  95.57  &  58.76   \\
                                       & Base+$L{\!}_{G{\!}R}$      &  89.49  &  61.10   \\
                                       & Base+$G{\!}_{M{\!}L}$+$S{\!}_{G{\!}R}$      &  91.77  &  55.38   \\
                                       & Base+$G{\!}_{M{\!}L}$+$L{\!}_{G{\!}R}$      &  94.48  &  55.15   \\
                                       & Base+$L{\!}_{G{\!}R}$+$S{\!}_{G{\!}R}$      &  90.85  &  56.44   \\
                                       & Ours (Base+MSAC)                       &  \textbf{85.29}  &  \textbf{62.95}  \\
        \midrule[\heavyrulewidth]
        \multirow{8}{*}{ODIN}          & Base                       &  95.41  &  59.08   \\
                                       & Base+$S{\!}_{G{\!}R}$      &  94.25  &  59.05   \\
                                       & Base+$G{\!}_{M{\!}L}$      &  95.11  &  59.14   \\
                                       & Base+$L{\!}_{G{\!}R}$      &  89.56  &  60.91   \\
                                       & Base+$G{\!}_{M{\!}L}$+$S{\!}_{G{\!}R}$      &  91.58  &  55.42   \\
                                       & Base+$G{\!}_{M{\!}L}$+$L{\!}_{G{\!}R}$      &  94.45  &  55.22   \\
                                       & Base+$L{\!}_{G{\!}R}$+$S{\!}_{G{\!}R}$      &  90.85  &  56.44   \\
                                       & Ours (Base+MSAC)                       &  \textbf{85.37} &  \textbf{62.92}   \\
        \midrule[\heavyrulewidth]
        \multirow{8}{*}{ReACT}         & Base                       &  94.38  &  56.31   \\
                                       & Base+$S{\!}_{G{\!}R}$      &  93.36  &  57.44   \\
                                       & Base+$G{\!}_{M{\!}L}$      &  96.17  &  58.49   \\
                                       & Base+$L{\!}_{G{\!}R}$      &  90.88  &  55.74   \\
                                       & Base+$G{\!}_{M{\!}L}$+$S{\!}_{G{\!}R}$      &  94.55  &  54.54   \\
                                       & Base+$G{\!}_{M{\!}L}$+$L{\!}_{G{\!}R}$      &  92.43  &  57.38   \\
                                       & Base+$L{\!}_{G{\!}R}$+$S{\!}_{G{\!}R}$      &  91.34  &  57.33   \\
                                       & Ours (Base+MSAC)                       &  \textbf{87.78}  &  \textbf{60.56}  \\
        \midrule[\heavyrulewidth]
        \multirow{8}{*}{Ma\_Dis}       & Base                       &  92.73  &  \textbf{61.62}   \\
                                       & Base+$S{\!}_{G{\!}R}$      &  95.48  &  53.92   \\
                                       & Base+$G{\!}_{M{\!}L}$      &  97.12  &  41.93   \\
                                       & Base+$L{\!}_{G{\!}R}$      &  93.25  &  48.47   \\
                                       & Base+$G{\!}_{M{\!}L}$+$S{\!}_{G{\!}R}$      &  93.19  &  49.94   \\
                                       & Base+$G{\!}_{M{\!}L}$+$L{\!}_{G{\!}R}$      &  94.11  &  51.37   \\
                                       & Base+$L{\!}_{G{\!}R}$+$S{\!}_{G{\!}R}$      &  94.37  &  56.01   \\
                                       & Ours (Base+MSAC)                       &  \textbf{91.55}  &  56.97   \\
        \bottomrule
    \end{tabular}
\end{table}

Similarly, it can be easily perceived that, for all employed OOD detection methods, our proposed MSAC approach enables the proposed base SER model to achieve the highest reliability performance.
Besides, compared with the existing four OOD detection methods, the proposed rODIN method attains the best reliability performance as well.
Furthermore, we find that when using additional speech attribute based gradient reversal strategy to optimize the proposed base model, the reliability performance generally outperforms that of the multi-task learning strategy-based approaches.
Consider the rODIN method, for instance.
Our proposed base SER model acquires the FPR95 and AUROC of 95.34\% and 59.10\%.
When adding one additional speech attribute control, it can be viewed that the $Base+G{\!}_{M{\!}L}$ performs even worse than the baseline, while the $Base+S{\!}_{G{\!}R}$ and $Base+L{\!}_{G{\!}R}$ achieve marginal enhancement over the baseline.
When exerting control over two additional speech attributes, the $Base+L{\!}_{G{\!}R}+S{\!}_{G{\!}R}$ outperforms $Base+G{\!}_{M{\!}L}+L{\!}_{G{\!}R}$ and $Base+G{\!}_{M{\!}L}+S{\!}_{G{\!}R}$, yet, they all exhibit varying degrees of decline in AUROC performance.
Comparatively, when incorporating the proposed MSAC, our proposed SER model obtains a 5.98\% reduction in FPR95 and a 4.02\% improvement in AUROC compared with the baseline.

\section{Conclusions And Future Work}
\label{sec:CONCLUSIONS}
In this paper, we introduce a novel unified SER workflow, MSAC-SERNet, based on the data distribution of various speech attributes for both single-corpus and cross-corpus SER. In addition, we pioneer an exploration into the reliability performance of the proposed MSAC-SERNet framework and propose a new OOD detection method for the SER tasks. 
Comprehensive experiments conducted on six public emotion corpora demonstrate that our proposed MSAC-SERNet not only achieves superior recognition and generalization results compared to state-of-the-art (SOTA) single-corpus and cross-corpus SER approaches but also consistently outperforms the baseline SER model in all aspects, including recognition, generalization, and reliability. 
Our findings highlight that intricately modeling speech emotion by exerting additional precise control over diverse speech attributes consistently enhances the overall performance of the proposed SER workflow.
Furthermore, we observe that commonly employed OOD detection techniques in CV and NLP fields, such as the input positive perturbation of the ODIN method, may not necessarily be applicable to the SER task. This implies the necessity for developing an OOD detection method more tailored to the inherent nature of SER tasks.

In future work, we will explore exerting control over a broader range of additional speech attributes and leverage the characteristics of SER tasks to propose more efficient OOD detection methods.

\iffalse
{\appendix[Proof of the Zonklar Equations]
Use $\backslash${\tt{appendix}} if you have a single appendix:
Do not use $\backslash${\tt{section}} anymore after $\backslash${\tt{appendix}}, only $\backslash${\tt{section*}}.
If you have multiple appendixes use $\backslash${\tt{appendices}} then use $\backslash${\tt{section}} to start each appendix.
You must declare a $\backslash${\tt{section}} before using any $\backslash${\tt{subsection}} or using $\backslash${\tt{label}} ($\backslash${\tt{appendices}} by itself
 starts a section numbered zero.)}
\fi

\bibliographystyle{ieee}
\bibliography{MSAC}

\newpage

\iffalse
    \section{Biography Section}
    If you have an EPS/PDF photo (graphicx package needed), extra braces are
     needed around the contents of the optional argument to biography to prevent
     the LaTeX parser from getting confused when it sees the complicated
     $\backslash${\tt{includegraphics}} command within an optional argument. (You can create
     your own custom macro containing the $\backslash${\tt{includegraphics}} command to make things
     simpler here.)
     
    \vspace{11pt}
    
    \bf{If you include a photo:}\vspace{-33pt}
    \begin{IEEEbiography}[{\includegraphics[width=1in,height=1.25in,clip,keepaspectratio]{fig1}}]{Michael Shell}
    Use $\backslash${\tt{begin\{IEEEbiography\}}} and then for the 1st argument use $\backslash${\tt{includegraphics}} to declare and link the author photo.
    Use the author name as the 3rd argument followed by the biography text.
    \end{IEEEbiography}
    
    \vspace{11pt}
    
    \bf{If you will not include a photo:}\vspace{-33pt}
    \begin{IEEEbiographynophoto}{John Doe}
    Use $\backslash${\tt{begin\{IEEEbiographynophoto\}}} and the author name as the argument followed by the biography text.
    \end{IEEEbiographynophoto}
    
    \vfill

\fi

\end{document}